\documentclass[cits]{PoS}

\usepackage{epsfig}
\usepackage{wrapfig}

\newcommand{\gev}{~\mathrm{GeV}}
\newcommand{\mev}{~\mathrm{MeV}}

\newcommand{\pp}{\partial}

\newcommand{\lk}{\left<}

\newcommand{\rk}{\right>}

\title{Chiral Aspects of Improved Staggered Fermions with 2+1-Flavors
  from the HotQCD Collaboration}

\ShortTitle{Chiral Aspects of Improved Staggered Fermions with 2+1-Flavors}

\author{\speaker{Wolfgang S\"oldner} { [for the HotQCD collaboration]}
  \thanks{ This work has been supported in part by contracts
    DE-AC02-98CH10886 and DE-FC02-06ER-41439 with the U.S.  Department
    of Energy and contract 0555397 with the National Science
    Foundation. The numerical calculations have been performed using
    USQCD resources at Fermilab and JLab, the BlueGene/L at the New
    York Center for Computational Sciences (NYCCS), and the BlueGene/L
    at the J\"ulich Supercomputing Center.}  \thanks{HotQCD
    Collaboration members are: A.~Bazavov, T.~Bhattacharya, M.~Cheng,
    N.H.~Christ, C.~DeTar, S.~Gottlieb, R.~Gupta, P.~Hegde,
    U.M.~Heller, C.~Jung, F.~Karsch, E.~Laermann, L.~Levkova, C.~Miao,
    R.D.~Mawhinney, S.~Mukherjee, P.~Petreczky, D.~Renfrew,
    C.~Schmidt, R.A.~Soltz, W.~S\"oldner, R.~Sugar,
    D.~Toussaint, W.~Unger, P.~Vranas}\\
  Helmholtz International Center for FAIR (HIC for FAIR),
  Max-von-Laue-Str. 1, D-60438 Frankfurt am Main, Germany\\
  E-mail: \email{w.soeldner@gsi.de}}

\abstract{We present recent results from lattice simulations of 2+1
  flavors of improved staggered fermions at zero baryon number density
  near the high temperature crossover. Included are new results from
  simulations of asqtad fermions at $N_\tau = 12$ and a nearly
  physical Goldstone pion mass and from simulations of HISQ fermions
  at $N_\tau = 6$ and 8.  We focus on observables sensitive to chiral
  symmetry and confinement. A companion HotQCD talk discusses the
  effects of staggered-fermion taste-symmetry breaking on
  thermodynamic quantities.}

\FullConference{The XXVIII International Symposium on Lattice Field
  Theory,
  Lattice2010\\
  June 14-19, 2010\\
  Villasimius, Italy}

\begin{document}

\begin{figure}[tp]
  \centerline{ \epsfig{file=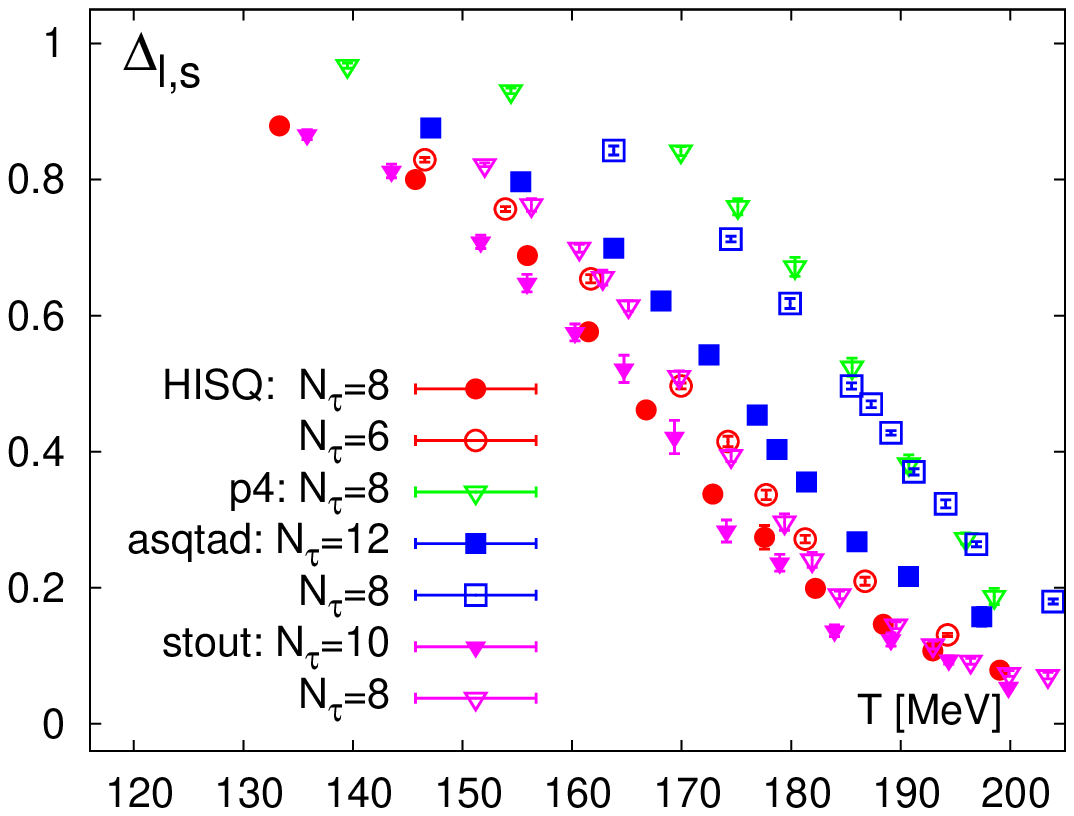, width = .43\textwidth}
    \epsfig{file=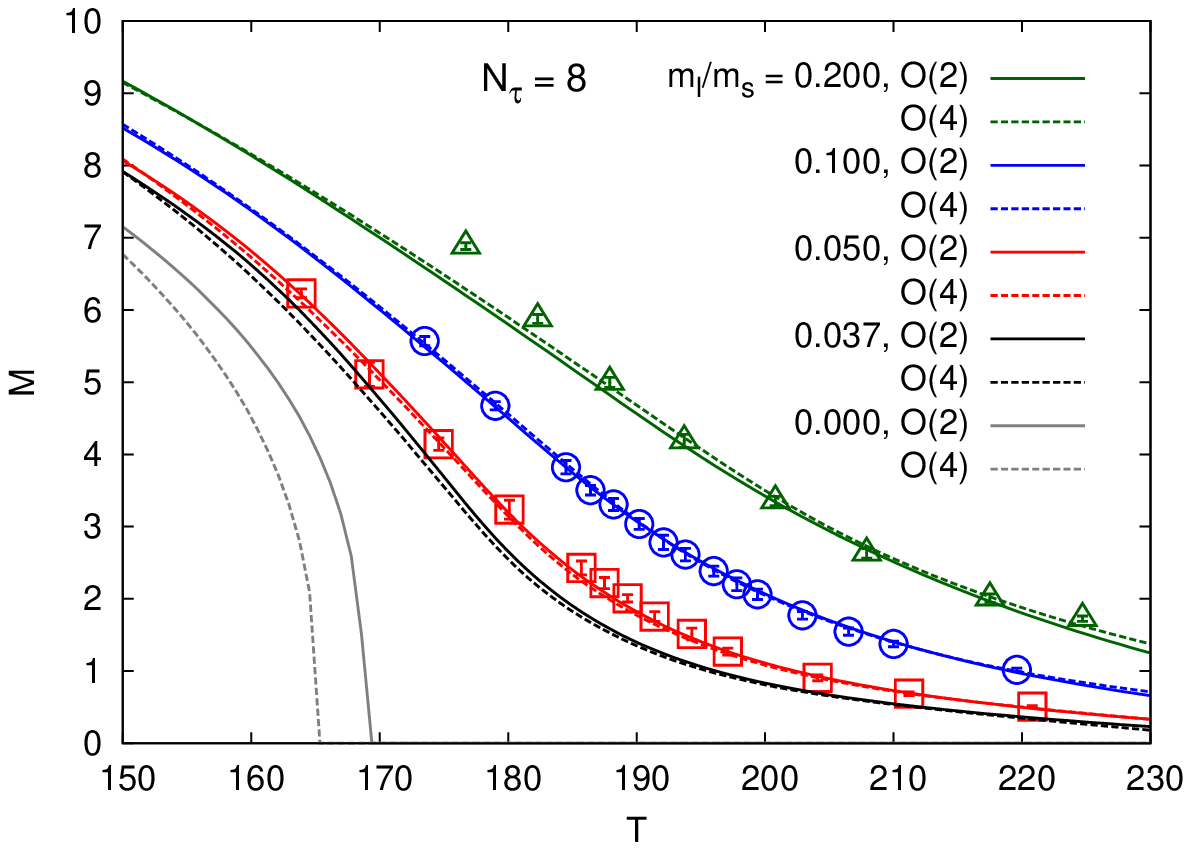, width = .46\textwidth}}
  \caption{\label{fig:chicond} Left: $\Delta_{l,s}$ as a function of
    temperature for HISQ/tree, asqtad, and p4 at different $N_\tau$ at
    light quark mass $m_l=0.05 m_s$. We compare with the stout
    action~\cite{stoutTc}. Right: Chiral condensate $M$ for the asqtad
    action parameterized by the scaling function $f_G$.}
\end{figure}

\section{Introduction}
Understanding the properties of QCD at finite temperature is important
for heavy ion collisions and the early universe.  Since for the most
part we need information about thermal QCD in the nonperturbative
regime, lattice calculations are required. In the past couple of years
we have developed a clearer understanding of the importance of cutoff
effects in some thermodynamic quantities, and, as a consequence, the
need for lattice calculations closer to the continuum limit and with
improved actions with intrinsically smaller cutoff effects
~\cite{hotqcd,stoutTc}. Here we report on the status of ongoing
calculations of the HotQCD collaboration with two light and one
strange quark using the rooted staggered fermion formalism. In
particular we present new results for the asqtad action at smaller
lattice spacing, {\it i.e.}~with temporal lattice extent $N_\tau=12$,
and new results for HISQ~\cite{hisq} fermions with a tree-level
improved gauge action (HISQ/tree)~\cite{baza2} at $N_\tau = 6$ and 8,
which has intrinsically smaller cutoff effects. The strange quark mass
$m_s$ is set to its physical value. For the light quark mass $m_l$ we
keep the ratio $\frac{m_l}{m_s}$ fixed, resulting, approximately, in
lines of constant physics over the temperature range of this
study~\cite{hotqcd}. The scale is set by using
$r_1=0.3106(8)(18)(4)\mathrm{fm}$~\cite{milc09}, for the calculations
with HISQ/tree action we use $r_0=0.469$fm to set the scale.  The
companion contribution \cite{lat10Baz} focuses on cutoff effects.  In
this contribution we investigate the chiral aspects of the finite
temperature transition in $2+1$-flavor QCD.

\section{Chiral aspects of the quark gluon plasma (QGP)}
The light quark chiral condensate is defined as a derivative of the
QCD partition function with respect to the light quark mass, $\lk
\bar\psi\psi \rk_l = \frac{T}{V} \frac{\pp \ln Z}{\pp m_l}$. It is an
order parameter of the QCD phase transition in the massless limit but
needs a multiplicative renormalization. For finite quark mass, it also
requires an additive renormalization.  The combination $\Delta_{l,s}$
of light and strange quark chiral condensates at finite and zero
temperature $T$ cancels the multiplicative renormalization constant
and the additive renormalization constant linear in the quark mass,
\begin{equation}
  \Delta_{l,s}=\frac{ \lk \bar\psi\psi
          \rk_{l,T} - \frac{\hat m_l}{\hat m_s} \lk \bar\psi\psi
          \rk_{s,T}}{ \lk \bar\psi\psi \rk_{l,0} - \frac{\hat
            m_l}{\hat m_s} \lk \bar\psi\psi \rk_{s,0}}.
\end{equation}
As can be seen on the left side of Fig.~\ref{fig:chicond}, the chiral
condensate $\Delta_{l,s}$ drops rapidly in the transition region. As
discussed in \cite{lat10Baz} we observe that cutoff effects,
particularly from taste-symmetry breaking, tend to shift the
transition region to higher temperatures. Those effects are less
pronounced for the HISQ/tree action. The transition region at
$m_l=0.05 m_s$ for the new $N_\tau=12$ asqtad data is around $T
\approx 170 \mev$ at this light quark mass, and for the new $N_\tau=8$
HISQ/tree data it is less.  Let us take a closer look at the
transition region. In this region physics is captured by the singular
part of the partition function.  For sufficiently small (light) quark
mass $m_l$ the order parameter is described by a universal scaling
function $f_G$ which depends on the critical exponents of the $O(2)$
($O(4)$) universality class, the external fields $m_l$ and $m_s$
through $H=\frac{m_l}{m_s}$, and the reduced temperature $\Delta
T=\frac{T-T_c}{T_c}$, see Ref.~\cite{scaling} for more details.
\begin{figure}[tp]
  \centerline{ \epsfig{file=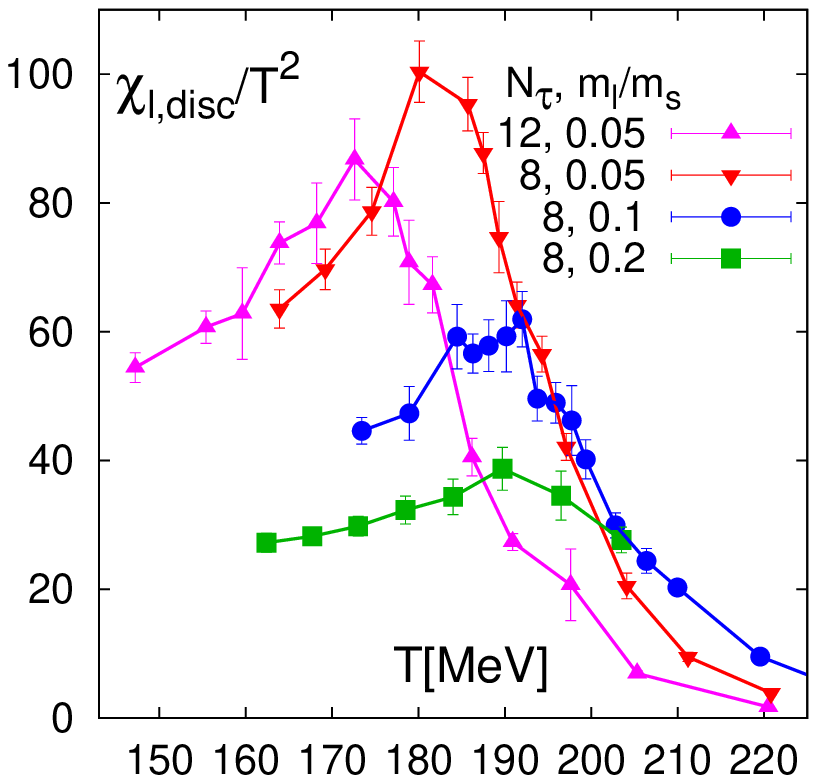,
      width = .55\textwidth}
    \epsfig{file=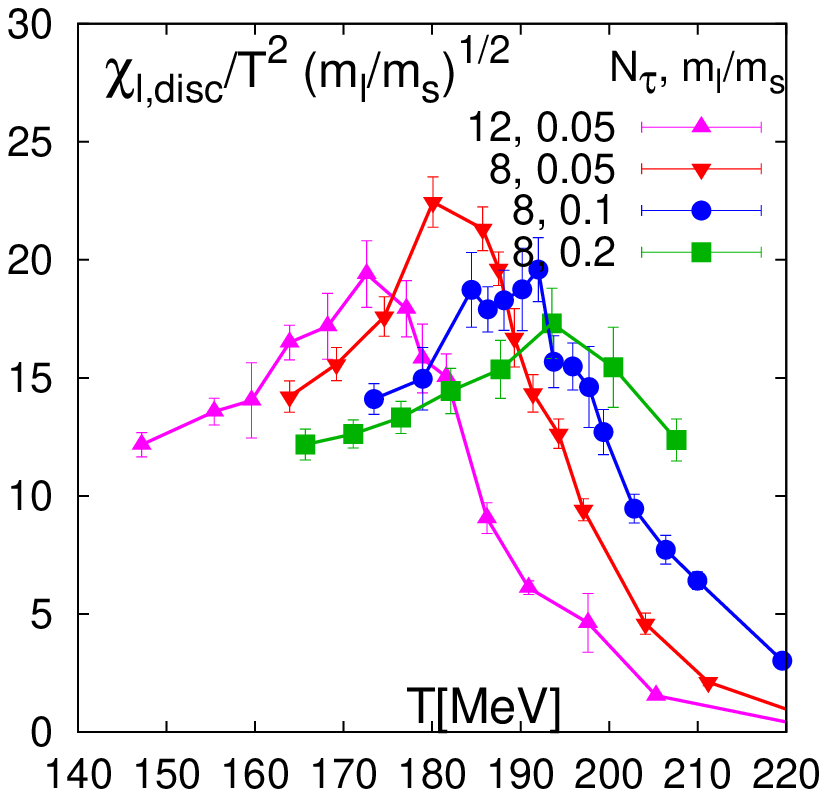, width =
      .55\textwidth}}
  \caption{\label{fig:chisusre}Disconnected chiral susceptibility for
    the asqtad action at different quark masses. On the right side
    $\chi_{\rm l, disc}$ has been rescaled by a factor
    $\sqrt{m_l/m_s}$.}
\end{figure}

On the right side of Fig.~\ref{fig:chicond} we plot the
multiplicatively renormalized chiral condensate $M_b
\equiv\frac{m_s}{T^4} \lk \bar\psi\psi \rk_l$ and its parameterization
given by the scaling function plus additional, scaling violating terms
which stem from the regular part of the partition function,
\begin{equation}
  M_b(T,m_l,m_s) = h^{1/\delta} f_G(t/h^{1/\beta\delta}) +
  a_t \Delta T H + b_1 H,
\end{equation}
where $a_t$ and $b_1$ are constants, $h=\frac{H}{h_0}$ and
$t=\frac{\Delta T}{t_0}$ with scales $h_0$ and $t_0$ which need to be
determined together with $T_c$ in the chiral limit~\cite{scaling}. The
constants $\delta$ and $\beta$ are critical exponents.  Fits based on
this parameterization yield good agreement with the data in the
transition region as well as for larger temperatures.  Less agreement
is found for larger quark masses and at lower temperatures.  The
latter observation may lead to the conclusion that the scaling window
in the hadronic phase is significantly smaller compared to the QGP
phase.

For a given lattice spacing the scaling analysis is especially useful
for defining the crossover temperature $T_p$ and for extrapolating
that temperature to its value at the physical light quark mass or at
the chiral limit of zero light quark mass ($T_c$).  The crossover can
be identified with the peak in the isosinglet chiral susceptibility,
{\it i.e.}, the peak in the derivative of the light quark chiral
condensate with respect to the light quark mass.  From the scaling
analysis we know explicitly how the crossover temperature depends on
the light quark mass as it approaches the critical point. We will
explore this approach in more detail in a future publication.
\begin{figure}[t]
  \centerline{
    \epsfig{file=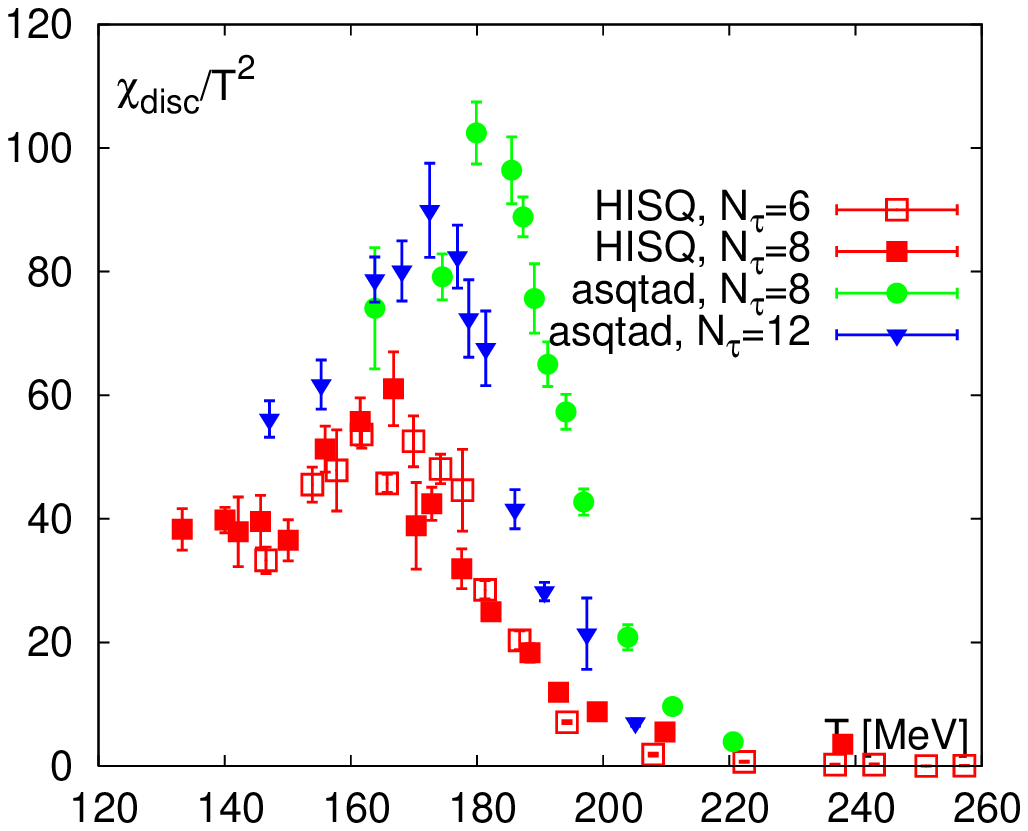, width = .49\textwidth}
    \epsfig{file=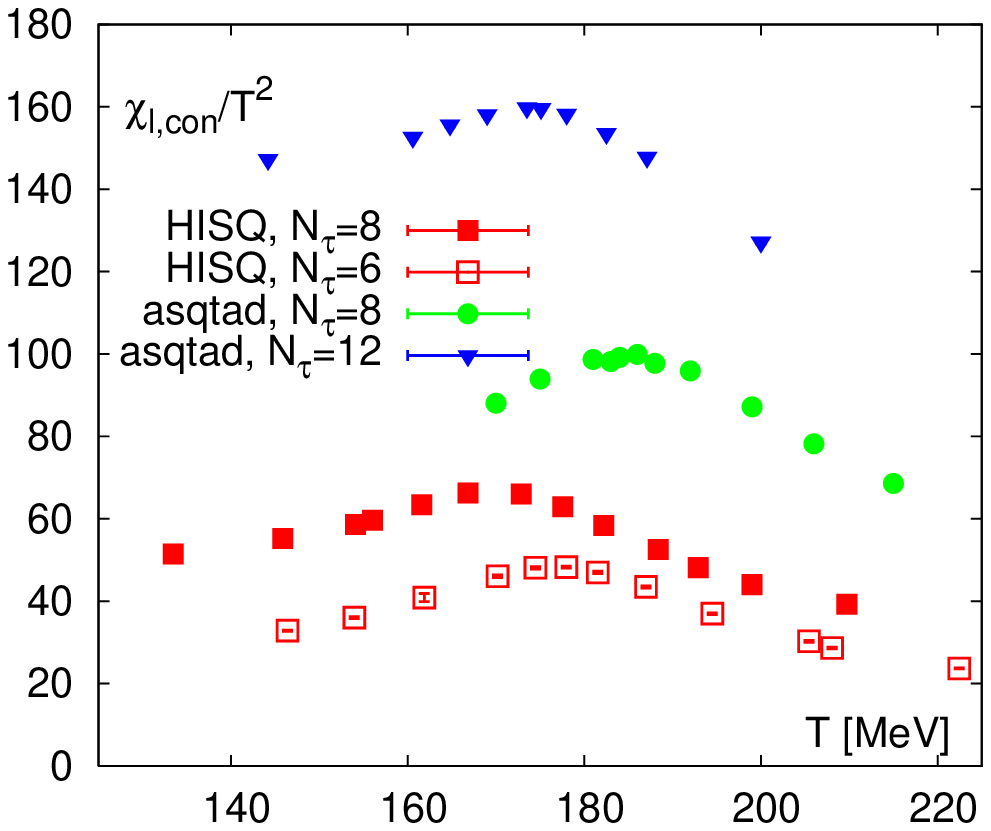, width = .49\textwidth}}
  \caption{\label{fig:chisus} Left: $\chi_{\rm l,disc}$ at fixed
    $m_l=0.05 m_s$ for different $N_\tau$ with HISQ/tree and asqtad
    action. Right: The connected chiral susceptibility.}
\end{figure}
For now we investigate the critical temperature $T_c$ by studying
peaks in the chiral susceptibility directly.  The ``full'' or
isosinglet chiral susceptibility $\chi_{\rm l}^{(n_f)}$ in terms of
the light one-flavor chiral condensate $\lk \bar\psi\psi \rk_l$ is
given by
\begin{eqnarray}
  &  \chi_{\rm l}^{(n_f)} \equiv 2 \frac{T}{V}  \frac{\partial}{\partial m_l}
   \lk \bar\psi\psi \rk_l 
  \equiv \chi_{\rm l,disc} + \chi_{\rm l,con} \quad \mathrm{with} \\
  &\chi_{\rm l,disc} = {T \over 4 V} \left\{
    \langle\bigl( {\rm Tr} D_l^{-1}\bigr)^2  \rangle -
    \langle {\rm Tr} D_l^{-1}\rangle^2 \right\}\;
  \quad \mathrm{and} \quad
  \chi_{\rm l,con} \equiv \frac{1}{2} \sum_x \left< D_l^{-1}(x,0)
    D_l^{-1}(0,x) \right>,
\end{eqnarray}
where $D_l$ is the light quark Dirac operator and $V$ is the spatial
volume.  We plot the disconnected chiral susceptibility $\chi_{\rm
  l,disc}$ for different $N_\tau$ and $m_l$, see left plot in
Fig.~\ref{fig:chisusre}. For all parameters the disconnected chiral
susceptibility develops a clear peak structure.  The height as well as
the position of the peak shows a sizable quark mass dependence. On the
left plot in Fig.~\ref{fig:chisus} we illustrate the cutoff dependence
of $\chi_{\rm l,disc}$. We find a mild dependence on the lattice
spacing in the height and position of the peak.  For the HISQ/tree
action these cutoff effects are reduced compared with the asqtad
action. The peak of $\chi_{\rm l,disc}$ at finite $N_\tau$ and $m_l$
defines pseudocritical temperatures $T_p$, which we use to determine
the critical temperature $T_c$ in the chiral and continuum limit.

Before we present this analysis we consider, briefly, the quark mass
dependence of $\chi_{\rm l, disc}$. For sufficiently small quark mass
the chiral condensate in the vicinity of the phase transition can be
understood in terms of the three-dimensional $O(N)$ model.  Below the
crossover temperature and for sufficiently small light quark mass the
quark mass dependence of the chiral condensate and its
susceptibilities is controlled by contributions from Goldstone modes.
These modes contribute a term proportional to $\sqrt{m_l}$ to the
chiral condensate which, in turn, leads to a $1/\sqrt{m_l}$ divergence
in the chiral susceptibility. These features are evident in
Fig.~\ref{fig:chisusre}.

In order to extract a continuum extrapolated critical temperature we
first determine the pseudocritical temperature for several $N_\tau$
and $m_l$ by fitting the peak position of $\chi_{\rm l,disc}$ with
different fitting ans\"atze where we allow for an asymmetric shape.
The different ans\"atze allow for an estimate of systematic error in
$T_p$.  The peak location in $\chi_{\rm l,disc}$ is controlled by the
singular part of the partition function. In the chiral limit, the
disconnected chiral susceptibility $\chi_{\rm l,disc}$ at the
pseudocritical temperature exhibits critical behavior, $\chi_{\rm
  l,disc}(T_p)\sim m_l^{\frac{1}{\delta}-1}$, with the critical
exponent $\delta$.  To extrapolate the pseudocritical temperature we
use an ansatz motivated by $O(N)$ models in leading order,
\begin{wrapfigure}[14]{tr}{.51\textwidth}
  \vspace{-5pt}
  \centerline{
 \epsfig{file=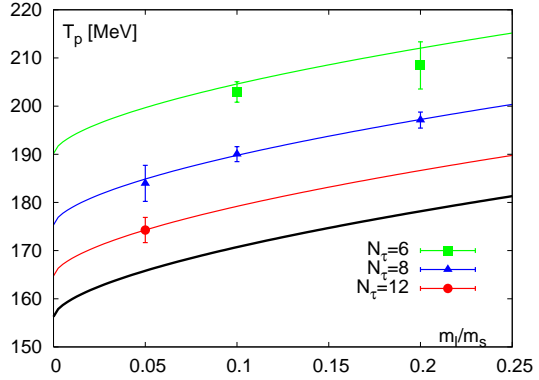, width = .5\textwidth}}
  \vspace{-13pt}
  \caption{\label{fig:Tc} The pseudocritical temperature for six
    ensembles at different $N_\tau$ and $m_l$. The colored curves
    correspond to the combined fit of the ansatz in
    Eq.~\protect\ref{eq:Tc}.  The black curve is the continuum
    extrapolation, $N_\tau \to \infty$.  }
\end{wrapfigure}
\begin{equation}
  T_p(m_l,N_\tau) =
  T_c +b \left( \frac{m_l}{m_s} \right)^d + c \frac{1}{N_\tau^2},
  \label{eq:Tc}
\vspace{5pt}
\end{equation}
where we use the critical exponent $d = \frac{1}{\beta \delta} \approx
0.54$ from $\mathrm{O}(N)$ model. The last term in Eq.~\ref{eq:Tc}
accounts for the $\mathcal{O}(a^2)$ cutoff dependence of staggered
fermions.  This ansatz is supposed to work well for sufficiently small
quark mass and lattice spacings. For the fitting we take into account
six values of $T_p$ obtained with the asqtad action at $N_\tau=6,8,12$
at different $m_l$. As can be seen in Fig.~\ref{fig:Tc}, the ansatz
works reasonably well and we obtain a stable fit when omitting $N_\tau
= 4$ data. The black curve in this plot gives the continuum
extrapolated $(N_\tau \to \infty)$ pseudocritical temperatures
obtained from the (combined) fit.  At physical mass parameters,
$\frac{m_l}{m_s} \simeq \frac{1}{27}$, we then obtain a continuum
extrapolated pseudocritical temperature for the chiral transition. Our
preliminary estimate for the transition temperature gives,
\begin{equation}
  T_p = \left( 164 \pm 6 \right) \mev.
\end{equation}
The error is a combined statistical and systematic estimate where also
the error of the scale setting coming from $r_1$ is included.

The connected chiral susceptibility $\chi_{\rm l,con}$ is an integral
over the scalar, flavor nonsinglet meson correlation function, and
thus is probing the thermal properties of the medium. Note that in the
continuum limit $\chi_{\rm l,con}$ can diverge in the thermodynamic
limit only if the ${\rm U}_A(1)$ symmetry is restored, which is not
expected to happen at the QCD transition temperature. In fact, lattice
calculations~\cite{screening} indicate that the scalar screening
masses develop a minimum at temperatures slightly above the transition
temperature.  Therefore, the connected chiral susceptibility is
expected to show a maximum above the chiral transition temperature,
even in the chiral limit. This is supported by our calculations by
comparing the peak position of the disconnected to the connected
chiral susceptibility, as shown in Fig.~\ref{fig:chisus}. Note that
$\chi_{\rm l,con}$ at finite quark mass suffers from additive and
multiplicative renormalization.  This issue will be addressed in more
detail in a future publication.

\vspace{-0.1cm}
\section{Deconfinement and chiral symmetry restoration}
\vspace{-0.1cm}

When discussing deconfinement one often investigates quantities like
the light~(l) and strange~(s) quark number susceptibility, $\chi_{l,s}
\equiv \frac{1}{V T} \frac{\partial^2 \ln Z}{\partial
  (\mu_{l,s}/T)^2}$, the energy density $\varepsilon$, or the Polyakov
loop. Beside the Polyakov loop, where a relation (if any) to the
singular part of the partition function is unknown, the other
quantities are also sensitive to critical behavior. However,
criticality is less pronounced in the sense that these quantities do
not diverge at $T_c$ in the chiral limit, in contrast, e.g., to the
disconnected chiral susceptibility.  For example, in the chiral limit
the temperature derivative of the quark number susceptibilities and
the trace anomaly in the vicinity of $T_c$ are respectively given by,
\begin{equation}
  \frac{\partial \chi_{l,s}}{\partial T} \sim c_r^{l,s} +A_{\pm}^{l,s} \left| 
    \frac{T-T_c}{T_c} \right|^{-\alpha} 
  \quad {\rm and} \quad
  (\varepsilon - 3p)/T^4 \sim a_r + b_r \frac{T-T_c}{T_c} + 
  c_\pm \left| \frac{T-T_c}{T_c} \right|^{1-\alpha}.
\label{eq:chiq}
\end{equation}
\begin{figure}[t]
  % \centerline{ \epsfig{file=chi2_s_6812_asqtad_large.pdf,
  \centerline{
 \epsfig{file=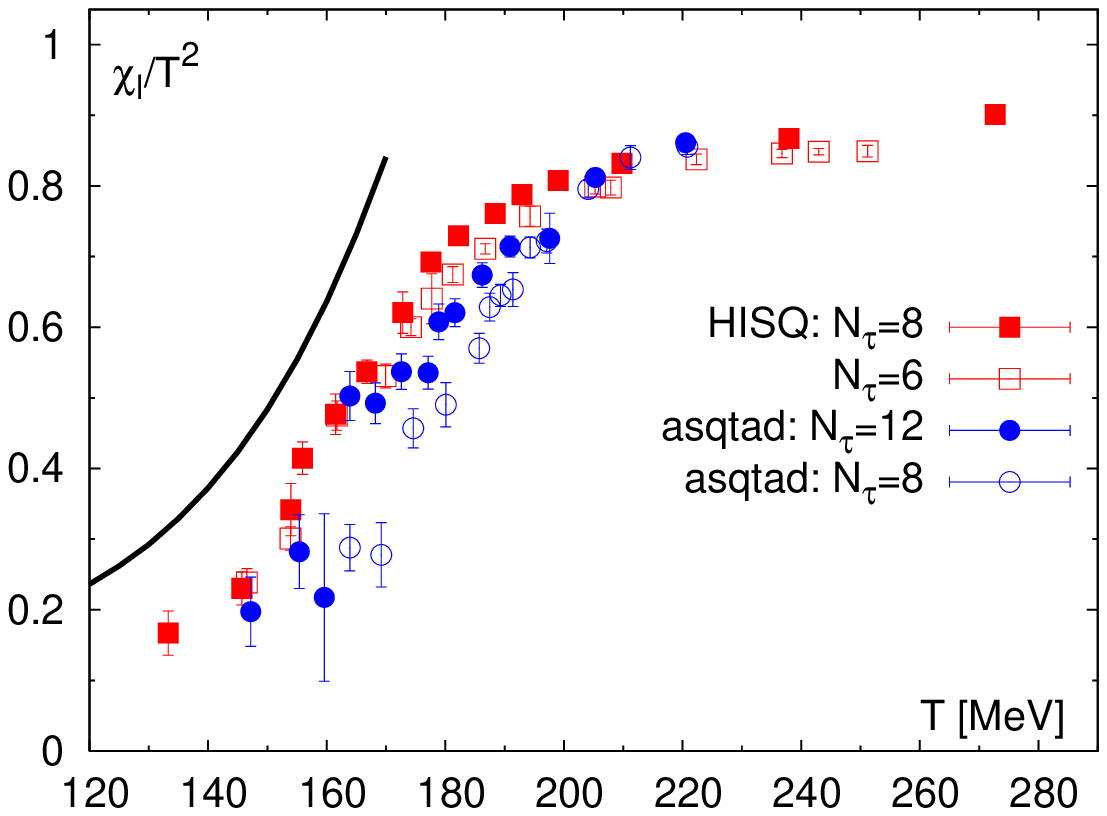, width = .49\textwidth}
 \epsfig{file=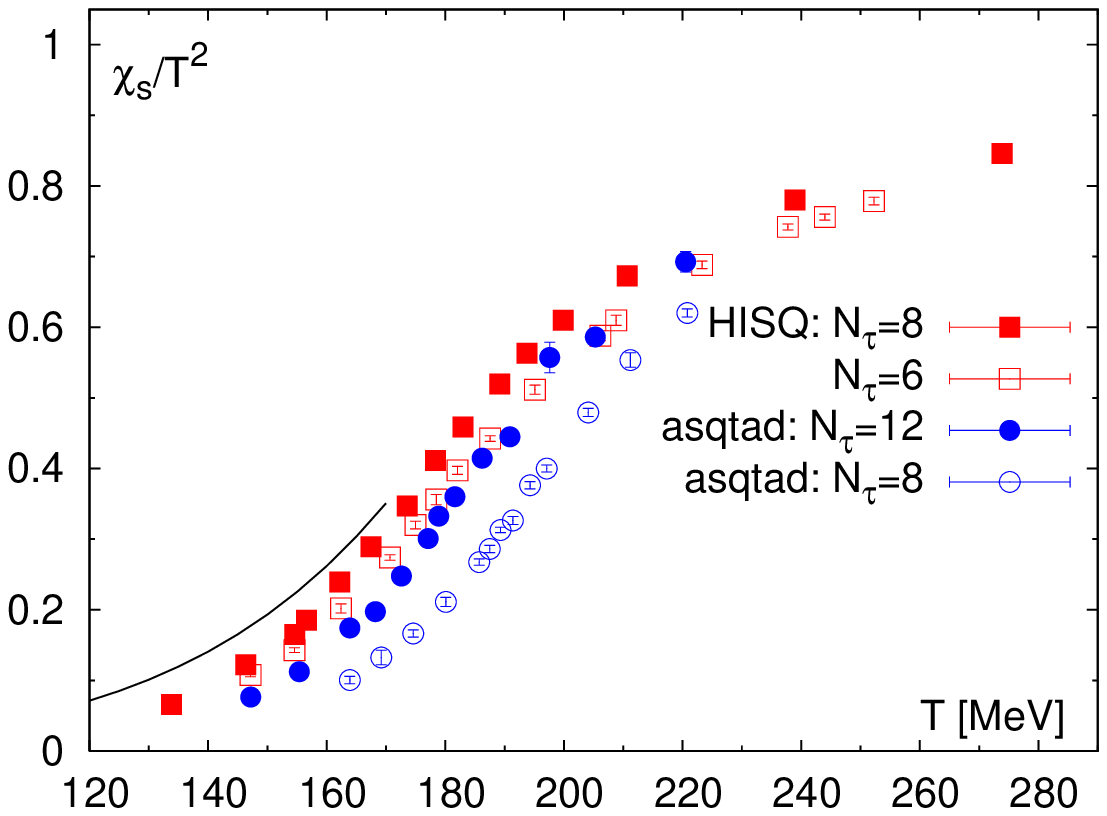, width = .49\textwidth}
}
\vspace{-5pt}
\caption{\label{fig:deconf} Light (left) and strange (right) quark
  number fluctuations for asqtad and HISQ/tree action compared to the
  HRG model (black curve).}
\vspace{-5pt}
\end{figure}
Note that at the critical temperature $T_c$ the slope is controlled by
a contribution that arises from the regular part of the free energy,
while its variation with temperature is given by the singular part.
Since the critical exponent $\alpha$ is negative it appears to be very
difficult to reliably extract information about $T_c$ from
$\chi_{l,s}$ because contributions from the regular part of the
partition function may hide critical behavior. High statistics lattice
data would be needed to discriminate the singular from the regular
contributions. Therefore, critical behavior in state-of-the-art
lattice calculations is best seen in quantities like the disconnected
chiral susceptibility.

Let us take a closer look at $\chi_l$ and $\chi_s$. For low
temperatures the relevant degrees of freedom which contribute to
$\chi_{l,s}$ are hadrons and $\chi_{l,s}$ is small. At high
temperatures the degrees of freedom are quarks and gluons.  The quark
number fluctuations are then related to the number of quarks and
increase with temperature towards the ideal gas limit.  In
Fig.~\ref{fig:deconf} we compare $\chi_l$ to $\chi_s$ for different
$N_\tau$ and actions.  We observe that in the vicinity of the
pseudocritical temperature $\chi_l$ as well as $\chi_s$ appear to be
rather smooth functions, the change in $\chi_l$, however, is more
pronounced. In terms of the above discussion this is related to the
smaller slope parameter $c_r^s$ compared to $c_r^l$. The relative
magnitude of these regular terms can quite naturally be understood in
terms of a hadron resonance gas (HRG) model.  A more detailed
investigation of this issue is currently ongoing. Note that our
calculations of deconfinement and chiral quantities indicate that
deconfinement and the chiral transition appear at about the same
temperature, see Fig.~\ref{fig:chisus} and Fig.~\ref{fig:deconf}.

\begin{figure}[tf]
  \centerline{ \epsfig{file=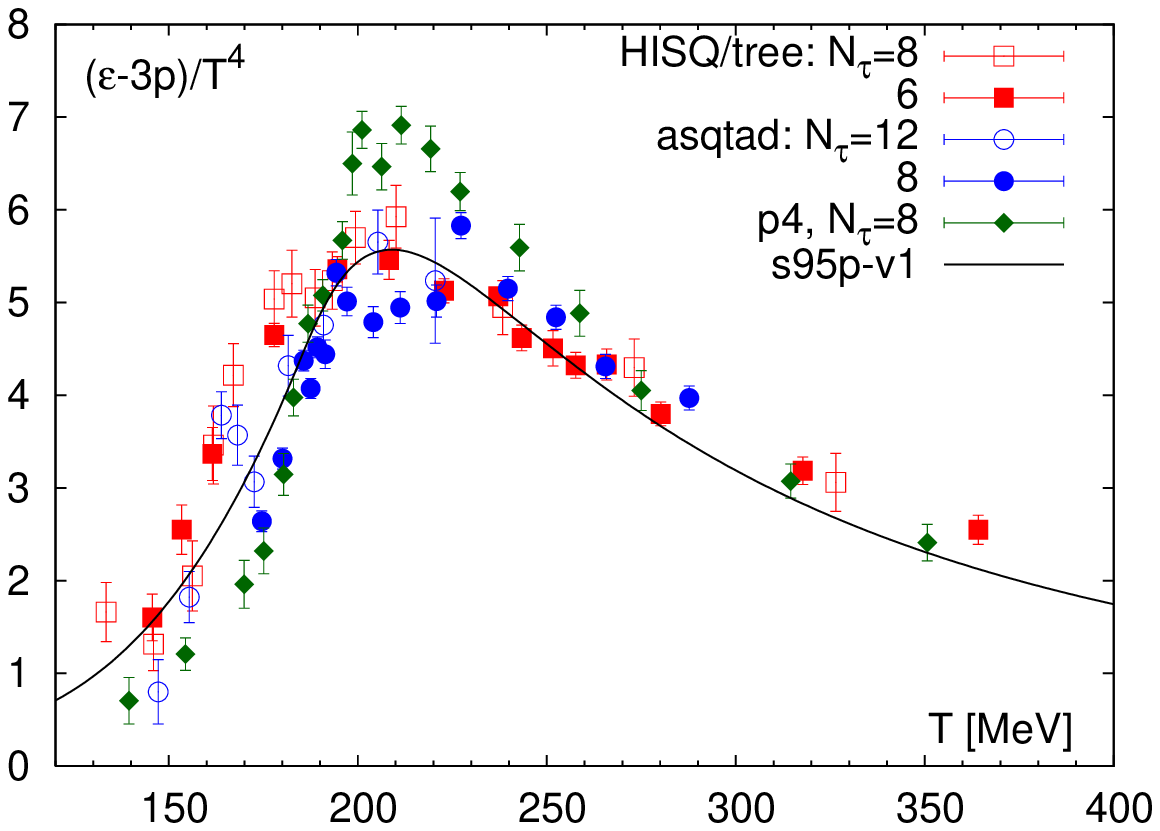, width = .48\textwidth}
    \epsfig{file=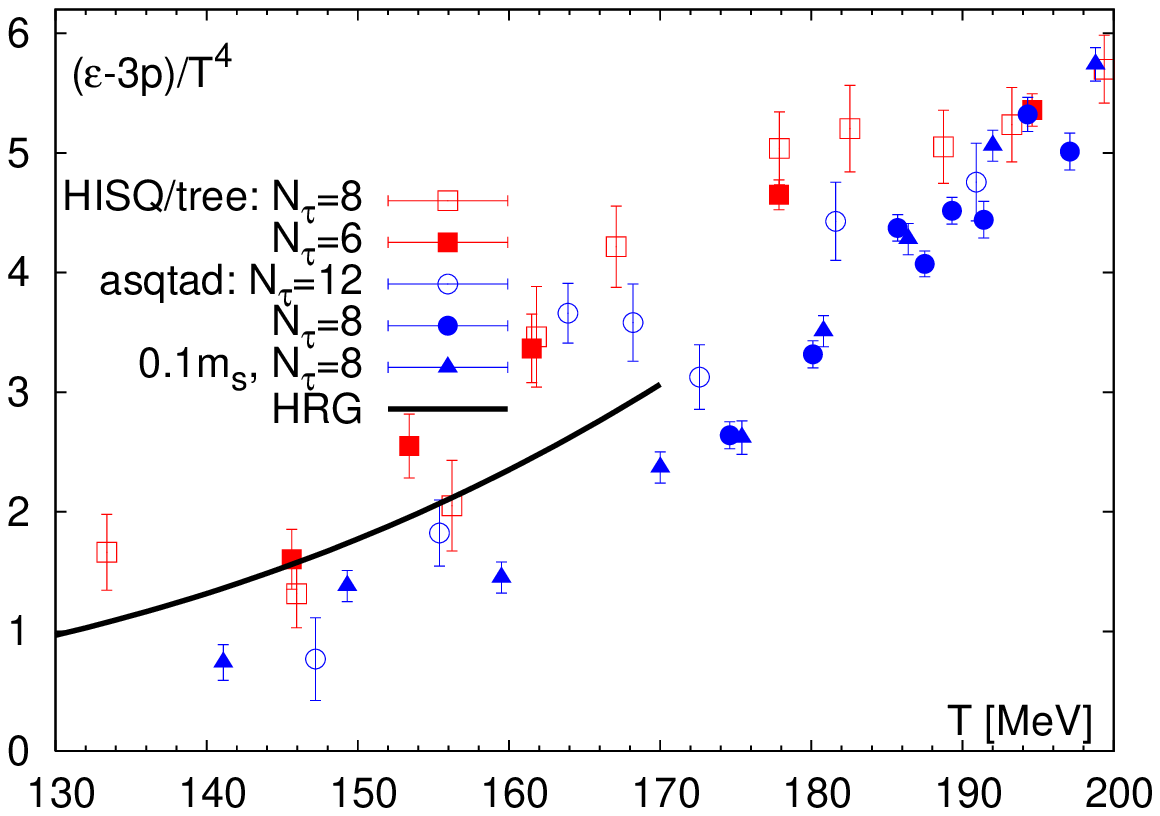, width = .48\textwidth}}
\vspace{-7pt}
  \caption{\label{fig:eos}Left: Trace anomaly for different actions at
    light quark mass $m_l=0.05m_s$ and the parameterization (black
    curve) from Ref.~\cite{pasi}. Right: Low temperature range of
    $(\varepsilon -3p)/T^4$ at $m_l=0.05m_s$ for HISQ/tree and asqtad
    action, and at $m_l=0.1m_s$ for asqtad action at $N_\tau=8$. We
    compare to the HRG model where resonances up to $2.5\gev$ are
    included .}
\vspace{-8pt}
\end{figure}

In Fig.~\ref{fig:eos} we show the trace anomaly for different actions
and quark masses. As can be seen in the left plot, at large
temperatures $(\varepsilon -3p)/T^4$ decreases towards zero. This is
expected because at high temperatures the conformal symmetry is
gradually restored. In the transition region and above $(\varepsilon
-3p)/T^4$ has its largest values. With respect to critical behavior
the trace anomaly and the quark number susceptibilities share the same
functional dependence, see Eq.~\ref{eq:chiq}. Critical behavior thus
is a subdominant feature in the trace anomaly, just like in the quark
number susceptibilities.  Therefore, it may not be surprising that
$(\varepsilon -3p)/T^4$ obtains its largest value at a temperature
above the pseudocritical temperature.  At temperatures below the
pseudocritical temperature the trace anomaly decreases and,
eventually, follows the prediction of the HRG model~\cite{hrg}. In the
plot on the right in Fig.~\ref{fig:eos} we show the low temperature
region of $(\varepsilon -3p)/T^4$ in more detail.  Compared with our
earlier calculations we observe that our new $N_\tau=12$ asqtad and
$N_\tau=8$ HISQ/tree data appear to be closer to the HRG curve.
However, it is also evident that at the current stage of our
calculations the data at low temperatures is not yet conclusive and
statistics for the new data sets have to be improved.

\vspace{-0.15cm}
\section*{Outlook}
\vspace{-0.15cm}
In this contribution we have given a status report of ongoing
calculations within the HotQCD collaboration focusing on chiral
properties at finite temperature for QCD with $2+1$ flavors. We
supplemented our earlier efforts with new calculations with asqtad
fermions at $N_\tau=12$ and $m_l=0.05m_s$, and with HISQ/tree fermions
at $N_\tau=6,8$ at $m_l=0.05m_s$. With this new asqtad calculation we
extracted a preliminary continuum extrapolated transition temperature
at the physical point, $T_p \simeq (164 \pm 6 )\mev$.  An improved
determination of $T_c$ using universal scaling behavior is under
investigation. The new HISQ/tree data in addition to upcoming
HISQ/tree $N_\tau=12$ are much closer to the continuum limit and will
enable us to further improve the continuum extrapolation. We remark
that from our calculations we observe that the chiral transition and
deconfinement appear at about the same temperature. Relating critical
behavior to observables which are dominantly sensitive to
deconfinement, {\it i.e.} the sudden change of degrees of freedom, is
more subtle.

\end{document}